\documentclass{svmult}

\usepackage{mathptmx}       
\usepackage{helvet}         
\usepackage{courier}        
\usepackage{type1cm}        
\usepackage{makeidx}         
\usepackage{graphicx}        
\usepackage{multicol}        
\usepackage[bottom]{footmisc}

\makeindex             

\usepackage{amssymb}
\usepackage{amsmath}
\usepackage{multirow}

\graphicspath{{figures/}}

\usepackage{caption}
\usepackage{subfig}

\usepackage{url}
\usepackage{hyperref}
\usepackage{tikz}
\usepackage{algorithm}
\usepackage{algpseudocode}

\usepackage{color}

\definecolor{mygreen}{rgb}{0,0.6,0}
\definecolor{mygray}{rgb}{0.5,0.5,0.5}
\definecolor{mymauve}{rgb}{0.58,0,0.82}

\usepackage{listings}
\begin{document}

\title*{Designing Complex Interplanetary Trajectories for the Global Trajectory Optimization Competitions}
\titlerunning{Designing Complex Interplanetary Trajectories for the GTOCs}

\author{Dario Izzo, Daniel Hennes, Lu{\'i}s F. Sim{\~o}es, and Marcus M{\"a}rtens}
\authorrunning{Dario Izzo et al.}

\institute{Dario Izzo \at European Space Agency, ESTEC, 2201AZ Noordwijk, The Netherlands, \\\email{dario.izzo@esa.int}
\and Daniel Hennes \at DFKI GmbH, 28359 Bremen, Germany \email{daniel.hennes@dfki.de}
\and Lu{\'i}s F. Sim{\~o}es \at Vrije Universiteit Amsterdam, 1081 HV Amsterdam, The Netherlands \\\email{luis.simoes@vu.nl}
\and Marcus M{\"a}rtens \at TU Delft, 2628 CD Delft, The Netherlands \email{m.maertens@tudelft.nl}}
%
%
\maketitle


\abstract{The design of interplanetary trajectories often involves a preliminary search for options later refined/assembled into one final trajectory. It is this broad search that, often being intractable, inspires the international event called Global Trajectory Optimization Competition. In the first part of this chapter, we introduce some fundamental problems of space flight mechanics, building blocks of any attempt to participate successfully in these competitions, and we describe the use of the open source software PyKEP to solve them. In the second part, we formulate an instance of a multiple asteroid rendezvous problem, related to the 7th edition of the competition, and we show step by step how to build a possible solution strategy. In doing so, we introduce two new techniques useful in the design of this particular mission type: the use of an asteroid phasing value and its surrogates and the efficient computation of asteroid clusters. We show how the basic building blocks, sided to these innovative ideas, allow designing an effective global search for possible trajectories.
}

\section{Introduction}
The design of interplanetary trajectories is a fundamental part of any future endeavour for the exploration of our solar system and beyond. Be it a sample return mission to Mars, the exploration of one of our gas giants, the first-time probing of objects in the Kuiper Belt, an asteroid deflection mission or the removal of dangerous orbiting debris, the complex interplay between the trajectory details and the mission objectives is what ultimately defines the overall mission value. The complexity and the diversity of interplanetary trajectories can be most immediately appreciated by looking at some remarkable examples such as the \mbox{SMART-1} \cite{smart1} transfer to Moon orbit, the Cassini tour of the Saturn system \cite{cassini}, or the Messenger \cite{messenger} interplanetary transfer to Mercury. Many interplanetary trajectories were successfully flown by past spacecraft and even more were designed in the process of learning how to best navigate around our solar system. An outstanding example is that of the international event known as the Global Trajectory Optimization Competition (GTOC). Initiated in 2006, and currently heading towards its 9th edition, the GTOC is an event where, as the official web portal \footnote{\url{http://sophia.estec.esa.int/gtoc_portal/}} reports: \emph {the best aerospace engineers and mathematicians world wide challenge themselves to solve a \lq\lq nearly-impossible\rq\rq\ problem of interplanetary trajectory design.} Providing a valid solution to these problems is a rather complex endeavour requiring a solid understanding of space-flight mechanics and a good dose of innovative thinking as all of the problems have unique characteristics and thus require the development of new methods and solution approaches built on top of available common knowledge. In this paper, we summarize part of the necessary (but often not sufficient) basic knowledge required to participate to these competitions and we report and discuss, as an example, part of the solution strategy we employed to design our submission to the 7th edition. We base the selection of techniques reported on past GTOC editions, mainly focused on the preliminary design of low-thrust missions neglecting effects of a third body gravitational attraction.

The paper is divided in two main sections as follows: in Sect. \ref{sec:bb} we describe fundamental problems that are often encountered during GTOCs. These include space flight mechanics problems (Sect. \ref{sec:sfm}), specific types of optimal control problems (Sect. \ref{sec:ocp}) and the efficient search of a computational tree (Sect. \ref{sec:ts}). In the second part of the chapter (Sect. \ref{sec:gtoc7}) we build a search strategy for multiple asteroid rendezvous in the main belt. We formally define the problem making large use of the 7th GTOC problem data in Sect. \ref{sec:problem_definition}. We then describe a set of new theoretical developments and their integration with the building blocks in a final algorithm described in Sect. \ref{sec:mobs} able to search for multiple asteroid rendezvous mission opportunities.

\section{Building blocks}
\label{sec:bb}
The design of a complex interplanetary mission is often made by (optimally) assembling solutions to a number of smaller problems, rather than tackling the problem as a whole. In this section, we introduce some of these basic \lq\lq building blocks,\rq\rq\ and we show the reader how to solve them on a computer using the open source code PyKEP \cite{pykep} \footnote{\url{https://github.com/esa/pykep/}}. PyKEP is an open source software library developed and maintained at the European Space Agency, which allows non-experts to perform research on interplanetary trajectory optimization. We report rough estimates on the CPU time employed to find solutions to these basic problems, assuming one single thread of an Intel(R) Core(TM) i7-4600U CPU having the clock at 3.3 GHz and with a cache of 4096 KB. Note that we use non dimensional units throughout the PyKEP examples, but any set of consistent units would also be compatible with PyKEP.

\subsection{Basic space flight mechanics problems}
\label{sec:sfm}
During the preliminary design of an interplanetary trajectory, and thus also in most GTOC problems, the spacecraft motion is approximated by that of a variable mass point, subject to the gravity attraction of one primary massive body with known gravitational parameter $\mu$, and to the spacecraft thrust $\mathbf T$. Denoting with $\mathbf r$, $\mathbf v$ and $m$ the position vector, velocity vector and mass of the spacecraft, the initial value (IV) problem describing its free motion in some inertial reference frame goes under the name of Kepler's problem (KP), mathematically defined as:
\begin{equation}
KP: \left\{
\begin{array}{l}
\ddot {\mathbf r} = - \frac{\mu}{r^3}\mathbf r \\
\mathbf r(t_s) = \mathbf r_s \\
\mathbf v(t_s) = \mathbf v_s \\
\end{array}
\right.
\end{equation}
where $t_s$ is the starting time and $\mathbf r_s,  \mathbf v_s$ the initial conditions. The position and velocity of a spacecraft at any time $t$ is then obtained by propagating the above equations. Numerical integration can be avoided in this well studied case by the use of the Lagrange coefficients technique (see \cite{battin, vallado} for implementation details). In PyKEP the KP is solved as follows:

\begin{lstlisting}
  from PyKEP import *
  rs = [1,0,0]; vs = [0,1,0]; t = pi / 2; mu = 1
  rf, vf = propagate_lagrangian(rs, vs, t, mu)
\end{lstlisting}

The CPU time requested by this operation is roughly constant across the whole spectrum of possible inputs.
Using the above mentioned hardware, we measured a mean time of roughly 40 microseconds per KP, corresponding to 250,000 KPs solved in one second.

The boundary value problem (BVP) associated with the free motion of our spacecraft is, instead, known as Lambert's Problem (LP) and is mathematically described as follows: 
\begin{equation}
LP: \left\{
\begin{array}{l}
\ddot {\mathbf r} = - \frac{\mu}{r^3}\mathbf r \\
\mathbf r(t_s) = \mathbf r_s \\
\mathbf r(t_f) = \mathbf r_f \\
\end{array}
\right.
\end{equation}
where $t_s$ is the starting time, $t_f$ the final time and $\mathbf r_s,  \mathbf r_f$ the boundary conditions. The search for techniques to efficiently solve this problem has an interesting history \cite{izzolambert}. The LP always results in at least one solution (the zero-revolutions solution) but, according to the value of $t_f - t_s$, may also result in several multi-revolutions solutions (mostly appearing in couples). In PyKEP (Python version) the LP is solved as follows:

\begin{lstlisting}
  from PyKEP import *
  rs = [1,0,0]; rf = [0,1,0]; t = 20 * pi / 2; mu = 1; mr = 5
  l = lambert_problem(rs, rf, t, mu, False, mr)
  v1 = l.get_v1()[0]
  v2 = l.get_v2()[0]
\end{lstlisting}

The CPU time requested by this operation depends on the number of existing multiple revolution solutions. If we limit ourselves to $mr=0$, that is to the zero-revolutions case (which indeed is often the most important), the algorithm implemented in PyKEP (Python version) can be considered to have constant CPU time across all possible inputs \cite{izzolambert}.
Using the above mentioned hardware, we measured a mean time of roughly 40 microseconds per LP, corresponding to 250,000 zero revolutions LPs solved in one second.

We then consider the spacecraft motion subject to a thrust force $\mathbf T$ constant in the inertial frame. This problem is also called the constant thrust problem (CTP). Since the spacecraft operates its propulsion system, some mass needs to be expelled in order to obtain an effect on the spacecraft acceleration. The efficiency of such a reaction process is described by the constant $I_{sp}$, i.e. the propulsion specific impulse. The corresponding initial value problem is:

\begin{equation}
CTP: \left\{
\begin{array}{l}
\ddot {\mathbf r} = - \frac{\mu}{r^3}\mathbf r + \frac{\mathbf T}{m}\\
\dot m = - \frac{T}{I_{sp}g_0} \\
\mathbf r(t_s) = \mathbf r_s, \mathbf v(t_s) = \mathbf v_s, m(t_s) = m_s 
\end{array}
\right.
\end{equation}
where $g_0$ is the Earth gravitational acceleration at sea level, typically set to $9.80665$ [m/s$^2$]. The technique we employ to efficiently solve this problem is a Taylor series numerical propagator \cite{taylor}. Other, more common, numerical propagators such as Runge-Kutta-Fehlberg would be significantly slower. In PyKEP (Python version) the CTP is solved, to a relative and absolute precision of $10^{-12}$ (see \cite{taylor} for the definition of such errors in the context of Taylor propagation), as follows:

\begin{lstlisting}
  from PyKEP import *
  rs = [1,0,0]; vs = [0,1,0]; ms = 10; t = 2 * pi 
  T = [0.01, 0.01, 0.01]; mu = 1; veff = 1
  rf, vf, mf = propagate_taylor(rs, vs, ms, T, t, mu, veff, -12, -12)
\end{lstlisting}

The CPU time requested by this operation depends linearly on the integration time $t$. Assuming the data in the example above (corresponding to one revolution along a circular orbit perturbed by a small thrust) and our reference thread performance, the algorithm implemented in PyKEP is able to solve the problem in roughly 230ms, corresponding to 45,000 CTP solved in one second.

\subsection{Optimal Control Problems}
\label{sec:ocp}
A fundamental aspect of interplanetary trajectory optimization problems where the spacecraft is equipped with a low-thrust propulsion system is the capability to solve Optimal Control Problems (OCPs) having the following mathematical description \cite{conway2010spacecraft}:
\begin{equation}
OCP: \left\{
\begin{array}{rl}
\mbox{find: } & \mathbf T(t) \in \mathcal F, \mathbf x_s, t_s, \mathbf x_f, t_f \\
\mbox{to minimize: } & J = \Phi(\mathbf x_s, t_s, \mathbf x_f, t_f) + \int_{t_s}^{t_f} \mathcal L(T(t), x(t), t) dt \\
\mbox{subject to: } & \\
        & \ddot {\mathbf r} = \frac{\mu}{r^3}\mathbf r + \frac{\mathbf T(t)}{m}\\
        & \dot m = - \frac{T(t)}{I_{sp}g_0} \\
        & \mathbf g(\mathbf x_s, t_s, \mathbf x_f, t_f) = \mathbf 0 \\
        & \boldsymbol \varphi(\mathbf x_s, t_s, \mathbf x_f, t_f) \le \mathbf 0
\end{array}
\right.
\end{equation}
where $\mathcal F$ is the functional space containing all piece-wise continuous functions, $\mathbf g$ are equality constraints and $\boldsymbol \varphi$ are inequality constraints. We also introduced the spacecraft state $\mathbf x = [\mathbf r, \mathbf v, m]$ to shorten our notation. Note that the above problem, and particularly some of its more complex variants, are still the subject of active research. In most cases the objective $J$ is one of 1) $J=t_f$ (time optimal control), 2) $J=m_f$ (mass optimal control), 3) $J= \int_{t_s}^{t_f} T^2(t) dt $ (quadratic control) or some combination of the above.

We will here shortly describe our approach (i.e. a direct approach based on the Sims-Flanagan transcription \cite{sims}) to solving the OCP as implemented in PyKEP, with the understanding that different approaches may perform better in some specific problems. Essentially, we divide the trajectory in $2n$ segments of constant duration $(t_f-t_s) / 2n$ and we consider the thrust $\mathbf T(t)$ as fixed along these segments in an inertial reference frame. The value of $\mathbf T$ fixed for each segment is denoted with $\mathbf T_i$. This allows the OCP to be transformed into an equivalent Non Linear Programming problem (NLP) \cite{NLP} having the following mathematical description:

\begin{equation}
NLP: \left\{
\begin{array}{rl}
\mbox{find: } & \mathbf T_i \in F, i = 1 .. 2n, \mathbf x_s, t_s, \mathbf x_f, t_f \\
\mbox{to minimize: } & J = \Phi(\mathbf x_s, t_s, \mathbf x_f, t_f) + \sum \int_{t_i}^{t_{i+1}} \mathcal L(\mathbf T_i, x(t), t) dt \\
\mbox{subject to: } & \mathbf x^- = \mathbf x^+ \\
        & \mathbf g(\mathbf x_s, t_s, \mathbf x_f, t_f) = \mathbf 0 \\
        & \boldsymbol \varphi(\mathbf x_s, t_s, \mathbf x_f, t_f) \le \mathbf 0
\end{array}
\right.
\end{equation}
where $F$ is a closed subset of $\mathbb R^3$, $\mathbf x^-$ is the spacecraft state as found propagating forward from $\mathbf x_s$ along the first $n$ segments, while  $\mathbf x^+$ is the spacecraft state as found propagating backward from $\mathbf x_f$ along the last $n$ segments. The equality constraint $\mathbf x^- = \mathbf x^+$ is called mismatch constraint, while the requirement $\mathbf T_i \in F$ is generally transformed into an inequality constraint called throttle constraint and representing a limit to the maximum thrust allowed by the spacecraft propulsion system.

As a fictitious example, we consider the transfer from Earth conditions to Mars conditions of a spacecraft having  $m_s=4500$ [kg] and $I_{sp} = 2500$ [s]. The spacecraft is equipped with a low-thrust propulsion system capable of thrusting at $T_{max} = 0.05$ [N]. We consider the minimum time problem, thus the following formal description:

\begin{equation}
EM: \left\{
\begin{array}{rl}
\mbox{find: } & t_s, t_f, m_f, T_{xi}, T_{yi}, T_{zi}, i = 1 .. 2n \\
\mbox{to minimize: } & J = t_f \\
\mbox{subject to: } &  \mathbf x^- = \mathbf x^+ \\
        & (T_{xi}^2 + T_{yi}^2 + T_{zi}^2)^2 \le T_{max}^2
\end{array}
\right.
\end{equation}
where $\mathbf x_s, \mathbf x_f$ are no longer in the decision vector as they are determined from $t_s$ and $t_f$ computing the Earth and Mars ephemerides. In PyKEP, the two constraints of the above EM problem are computed as follows, assuming the decision vector is known:

\begin{lstlisting}
  from PyKEP import *
  # Example Decision Vector for 10 segments
  n_seg = 10
  ts = epoch(0, 'mjd2000'); tf = epoch(350, 'mjd2000'); mf = 2400
  throttles = [0, 0, 1] * n_seg
  # Computing the planets positions and velocity (ephemerides)
  earth = planet.jpl_lp('earth')
  mars = planet.jpl_lp('mars')
  # Computing the equality and inequality constraints
  sc = sims_flanagan.spacecraft(4500, 0.05, 2500)
  rs, vs = earth.eph(ts)
  rf, vf = mars.eph(tf)
  xs = sims_flanagan.sc_state(rs, vs, sc.mass)
  xf = sims_flanagan.sc_state(rf, vf, mf)
  leg = sims_flanagan.leg(ts, xs, throttles, tf, xf, sc, MU_SUN)
  ceq = leg.mismatch_constraints()
  cineq = leg.throttles_constraints()
\end{lstlisting}

The CPU time requested for computing these constraints is $2n$ times that requested by the underlying CTP. Once equality, inequality constraints and the objective function computations are available, they can be used by an NLP solver to find the optimal solution. Widely spread solvers like IPOPT \cite{ipopt}, SNOPT \cite{snopt} and WORHP \cite{worhp} are an obvious choice and have indeed been successfully used in connection to this type of NLPs and more in general in GTOCs and interplanetary trajectory design.

\subsection{Tree Searches}
\label{sec:ts}

The various problems described in Sects. \ref{sec:sfm} and \ref{sec:ocp} can be used in the design of interplanetary trajectories, such as those of the GTOC problems, as building blocks of a more complex search strategy. Such a search strategy is often some form of tree search, where each node represents a trajectory that can be incrementally built towards the mission goal expanding one of its branches (e.g. adding a fly-by, a rendezvous or, more generically, a trajectory leg). The exact detail of the nodes definition and their possible branching must be carefully designed according to the problem under consideration. A concrete example is given in a later section to clarify such a process in one particular, selected case. For the time being, one may picture each node as representing a partial trajectory and the branching as the process to add one or more phases to such a trajectory. Branching involves the solution of one or more sub-problems such as an LP, or an OCP, etc. and its complexity may vary greatly. Due to the complexity of the search, it is often impossible to exhaustively search the entire tree of possibilities. Simple text book implementations of breadth first search (BFS) or depth first search (DFS) are exhaustive search-strategies that will eventually branch out every possible node, which is most of the time never an option due to the enormous tree size.

Consequently, one has to develop a strategy that explores only areas of the tree that give best results while staying within a reasonable computational budget, i.e. the number of sub-problems to be solved. The key aspect in the design of a tree search strategy is then, given a set of active leaf nodes (i.e. partial trajectories), to choose which one is worth branching and what branches are worth computing. Since each node represents only a partial interplanetary trajectory, its value with respect to the achievement of the final mission goal is not necessarily available. In a typical example, at each node the remaining propellant mass $r_m$ and the remaining mission time $r_t$ are known and, only in some cases, a partial objective value $J$ measuring the mission value achieved so far is available. Using this knowledge to decide what node to branch next is, as mentioned, of paramount importance.

\begin{figure}
\centering
\subfloat[Breadth-first-search (BFS)]{
    \includegraphics[width=0.325\columnwidth]{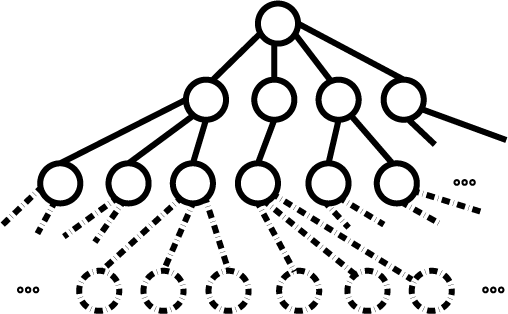}}
\subfloat[Depth-first-search (DFS)]{
    \includegraphics[width=0.325\columnwidth]{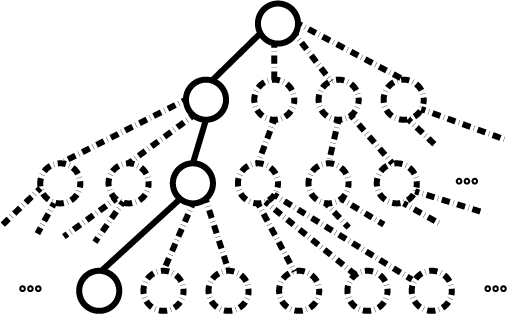}}
\subfloat[Beam-search (BS)]{
    \includegraphics[width=0.325\columnwidth]{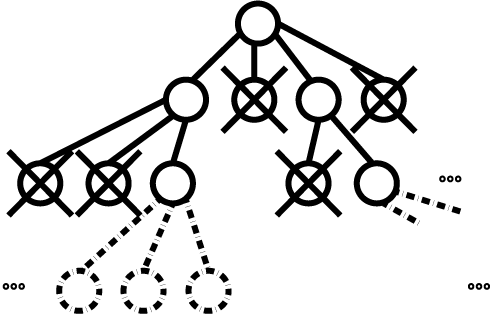}}
\caption{Different tree search strategies in comparison. Dotted nodes are yet to be explored. Crossed out nodes are pruned and will not be branched.}
\label{fig:treesearches}
\end{figure}
The text book implementation of DFS \cite{cormen2001c} can be improved by introducing a pruning criterion preventing nodes to be further expanded. Such a criteria can make use of $r_m$, $r_t$ and, when available, $J$ as well as of the information on the best full trajectory found so far which will be available rather soon during the search since the tree is searched in depth. The main problem with this approach is that its running time is very sensitive to the pruning criteria, but it cannot be estimated upfront. As a consequence when a tree search starts one needs to wait for it to finish in order to decide whether the pruning criteria was too strict or loose. During the 5th GTOC this strategy was used \cite{gtoc5} to explore the tree of Lambert's solutions that would then be converted into a low-thrust trajectory.

The text book implementation of BFS \cite{cormen2001c} can be improved by considering only a fixed number of nodes for branching at each depth. The nodes are prioritized using $r_m$, $r_t$ and, when available, $J$. The resulting tree search is a standard tree search called beam search (BS). A version of this tree search strategy was employed by Jet Propulsion Laboratory in the design of their winning trajectory in the 5th GTOC edition \cite{gtoc5jpl}. Figure~\ref{fig:treesearches} gives a schematic comparison of beam search with BFS and DFS. A multi-objective version of beam search was also used during the 7th GTOC by our team, resulting in a search strategy we called MOBS and that is described in detail in a later section of this paper. An advantage of the BS / MOBS approach is that the complexity to explore an additional tree depth is easily computed as the complexity of the sub-problem to solve times the beam size. It is thus possible to estimate rather accurately the running time of the search before starting it.

In some problems, it is not possible to make a fair selection among nodes having equal depth in the tree. In fact, in most problems, the tree depth information is not directly related to relevant physical phenomena and it is just an artifact of how the problem under consideration is mapped into a tree search problem. A fairer comparison can be made among nodes representing trajectories that have a similar remaining mission time $r_t$ at disposal to achieve their objectives. A tree search based on this simple idea, called Lazy Race Tree Search (LRTS) was used during the 6th GTOC and the resulting search strategy, employing self-adaptive differential evolution in the trajectory branching, received the gold ``Humies" award for human-competitive results produced by genetic and evolutionary computation \cite{gtoc6}.

A different approach to tree searches, and one that is most popular between AI practitioners as it proved to be able to deal with the vast combinatorial complexity of board games such as the game Go \cite{chaslot2006monte}, is the Monte Carlo Tree Search (MCTS). An implementation of the MCTS paradigm in the design of complex interplanetary trajectories was recently studied in the context of purely ballistic trajectories and fly-by sequences generation \cite{mcts} suggesting that its use may be competitive with beam search.

\section{Example: multiple-asteroid rendezvous mission}
\label{sec:gtoc7}
In this section we define a multiple-asteroid rendezvous mission. We reuse large part of the problem description released for the 7th edition of the GTOC and we describe a possible solution strategy. Our focus is on showing how new innovative ideas and design methods have to be developed and used aside the basic building blocks to allow for an efficient search of design options in this particular case, highlighting how the problem of \lq\lq interplanetary trajectory design\rq\rq\ is still far from being automated in its most general case.

\subsection{Problem definition}
\label{sec:problem_definition}
Consider a spacecraft $S$ having an initial mass $m_0 = m_s + m_p$, where $m_s$ is the dry mass and $m_p$ the propellant mass.  The spacecraft has a propulsion system defined by a maximum thrust $\tau_M$, and a specific impulse $I_{sp}$. The maximum acceleration allowed by the propulsion system is denoted by $\alpha_M = \tau_M / m_s$. The set $\mathcal A$ contains $N$ possible target asteroids of which the ephemerides (e.g. position and velocity) at each epoch $t_0 \in [\overline t_0, \underline t_0]$ are known or computable. We want to perform a preliminary search for possible multiple rendezvous missions allowing for the spacecraft to visit the largest possible number of asteroids within a maximum mission duration $tof$ and allowing for a minimum stay time $t_w$ on each of its visited asteroids. A visit is defined, mathematically, as a perfect match between the asteroid and the spacecraft positions and velocities. We focus on the case where the cardinality $N$ of the set $\mathcal A$ (i.e. the number of possible target asteroids) is in the order of thousands and we assume the spacecraft can be delivered on a chosen starting asteroid at a chosen starting date. 

This problem is relevant to the design of advanced asteroid belt exploration missions, such as the one considered in the 7th edition of the Global Trajectory Optimization Competition \cite{gtoc7:problem}, advanced In Situ Resource Utilization (ISRU) missions or future concepts such as the APIES concept \cite{d2006apies}, as well as to the design of multiple active debris removal missions (in which case the set $\mathcal A$ contains orbiting debris rather than asteroids). The asteroid belt scenario types are studied in depth here, but the novel methods proposed are of more general significance. As data set, we use the 16,256 asteroids from the main belt that were used during the GTOC7 competition. The ephemerides of such asteroids are computed from the orbital parameters assuming perfectly Keplerian orbits. The actual orbital parameters can be downloaded from the GTOC7 problem data at \url{http://sophia.estec.esa.int/gtoc_portal/}. Each asteroid of such a data set is assigned a consecutive index, so that one can write $A_j$, with $j \in [1, 16256]$ to identify uniquely the asteroid.

\begin{figure}
\centering
\includegraphics[width=0.8\columnwidth]{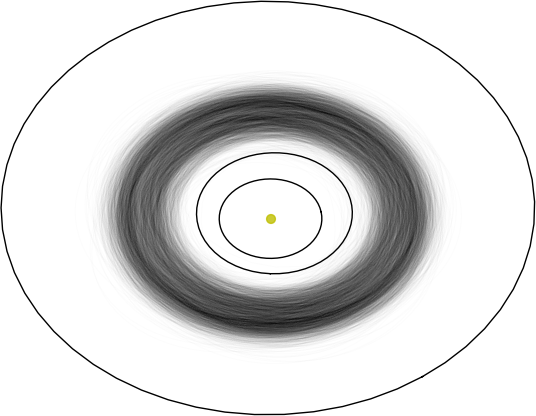}
\caption{Visualization of the orbits of 16,256 main belt asteroids considered for the GTOC7 problem. Going outward, the orbits of the Earth, Jupiter and Mars are also shown as reference.}
\label{fig:main_belt}
\end{figure}

\subsection{Asteroid phasing}
A fundamental problem in a multiple rendezvous mission is that of assessing which good transfer opportunities (target body, arrival epoch, arrival mass, etc ... ) are presented to a spacecraft $S$. Since it is computationally demanding to define and solve an optimal control problem (OCP) for each possible target body and launch / arrival window, we introduce a new quantity, easily computed and related to the cost of performing a given transfer: the phasing value, or asteroid phasing value in our chosen context. When good transfer opportunities from $A_s$ to $A_f$ exist at some epoch we say that those two asteroids are well \lq\lq phased\rq\rq\ and their phasing value will be small. Before introducing the formal definition of the phasing value, it is worth noting immediately how such a notion depends also on the spacecraft $S$ and its propulsion system and not only on the starting and final body.

\begin{figure}
\centering
\includegraphics[width=0.95\columnwidth]{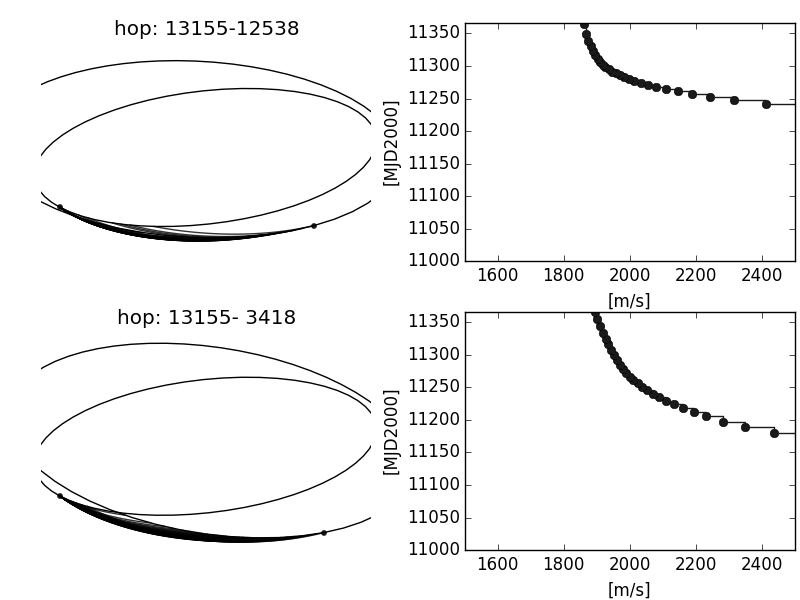}
\caption{Visual representation of the phasing value. Trajectories (left), Pareto front (right)}
\label{fig:phasing example}
\end{figure}

Assume the spacecraft $S$ to be on the asteroid $A_s$ at $t_0$ and consider possible Lambert problems (LP) to target the asteroid $A_f$. Let the starting ($t_s$) and final ($t_f$)  epochs vary freely in $[t_0, t_0 + T_M]$ and consider the minimization of two final objectives $\Delta V$ and $t_f$ only for trajectories for which the thruster can actually deliver the requested velocity increment, that is if $\Delta V \le \alpha_M \Delta T$, where $\alpha_M$ is the maximum value for the thruster acceleration and $\Delta T = t_f - t_s$. The $\Delta V$ is computed from the solution of the LP as $\Delta V = \Delta V_1 + \Delta V_2 $, where the two velocity increments represent the departure and arrival relative velocities along the Lambert solution. This results in the following two dimensional, two objectives, constrained optimization problem:
\begin{equation}
\begin{array}{rl}
\mbox{find: }  & t_s, t_f \in [t_0, t_0 + T_M]  \\
\mbox{to minimize: }  & f_1 = \Delta V, f_2 = t_f  \\
\mbox{subject to:} & \Delta V \le \alpha_M \Delta T \\ 
                   & t_s < t_f \\
\end{array}
\label{eq:define_phi}
\end{equation}
The quality of its Pareto front is proposed as a quantitative measure for the notion of phasing, henceforth referred to as \emph{phasing value} and indicated with
$\varphi(A_s, A_f)$, shortened from $\varphi (A_s, A_f, t_0, T_M, \alpha_M)$.
Note that one of the two objectives is the arrival epoch $t_f$ and not the total time of flight which is, in this case, not relevant, also note that $t_s$ indicates the starting epoch of the Lambert transfer and is not necessarily equal to $t_0$. As an example, take $A_s = A_{13155}$, $t_0 = 11000$ [MJD2000] and $\Delta T = 365.25$ [days] and solve the above problem for $A_f = A_{12538}$ and $A_f = A_{3418}$ and a value of $\alpha_M = 0.375$ $10^{-4}$ [m/s$^2$]. The resulting Pareto fronts (computed using MOEA/D \cite{zhang2007moea} and accounting for the constraints using a death penalty method \cite{bdack1991survey}) are shown in Figure \ref{fig:phasing example} together with all the transfer orbits in the front. In this case, one may state that $A_{3418}$ is better phased, at $t_0$, than $A_{12538}$  with respect to $A_{13155}$ (w.r.t. $T_M$ and $\alpha_M$). This definition has a straight forward application to the planning of multiple asteroid rendezvous missions as it allows to introduce a strict ordering over the set of possible target asteroids. In other words, given $A_s$, $t_0$ and $\Delta T$ and $\alpha_M$, one can rank all the possible target asteroids with respect to $\varphi$ and thus select the best for a further more detailed analysis.

The formal definition of Pareto front quality remains to be introduced. For the purpose of this work, the hypervolume \cite{zitzler1998multiobjective} is used. In our simple two-dimensional case the hypervolume can be quickly visualized as the area between the front and the vertical and horizontal line passing through a reference point. In Figure \ref{fig:phasing example} (graphs on the right) this is easily done as the reference point also corresponds to the maximum values of the axes. When using the hypervolume as a quality indicator for Pareto fronts, one must take care to select a reference point $p^*$. We may use the following reference point: $p^* = [\Delta T \tau_M / m_s, t_0+\Delta T]$. This definition ensures that whenever a feasible trajectory exists, its objectives are below the reference point. Note that the hypervolume has, in our case, the dimensions of a length and that larger values indicate better phasing values. In the example introduced, the computation of such a metric returns $\varphi = 1.1$ [AU] for the case $A_f = A_{12538}$ and $\varphi = 1.37$ [AU] for $A_f = A_{3418}$ indicating quantitatively that $A_f = A_{3418}$ is better phased.

\subsubsection{Phasing indicators}
The computation of the phasing value $\varphi$ is done referring to its definition given by Eq.(\ref{eq:define_phi}). A multi-objective optimization problem is solved and the hypervolume of the resulting Pareto front computed. While this procedure is fast in a single case, whenever a large number of phasing values are to be computed it does require significant computational resources. Since $\varphi$ is ultimately used to rank transfer opportunities, one may consider to compute different quantities and study their correlation to the ground truth defined by $\varphi$. Such an approach will only be valuable if the new quantities, which we refer to as phasing indicators, are computed with less computational cost with respect to $\varphi$ and the derived ranks have a high degree of correlation with those computed from $\varphi$. Two different phasing indicators are proposed and studied: the Euclidean indicator $d_e$ and the orbital indicator $d_o$.

The \emph{Euclidean indicator} is defined as $d_e = |\mathbf x_2 - \mathbf x_1|$, where $\mathbf x = [\mathbf r, \mathbf v]$, and contains information on both the asteroids relative positions and their relative velocities. The basic idea is that asteroids physically near to each other (and having a small relative velocity) are likely to be good candidates for an orbital transfer. The euclidean distance indicator can also be written as $d_e = \sqrt{|\Delta \mathbf r|^2 + |\Delta \mathbf v|^2}$ where $\Delta \mathbf r$ and $\Delta \mathbf v$ are the differences between the asteroid ephemerides. The main drawback of this indicator is that it is unable to distinguish between a case where the relative velocity eventually brings the asteroids closer and a case (e.g. having an identical $|\mathbf x_2 - \mathbf x_1|$) where the relative velocity tends to separate the asteroids. 

A different indicator, which we call the \emph{orbital indicator}, derives from the following simple linear model of an orbital transfer. Consider three points $P_0, P_1$ and $P$ undergoing a uniform rectilinear motion. These represent the two asteroids and the spacecraft. Assume that the motion of the three points is determined by the equations:
$$
\begin{array}{c}
\mathbf r_0 = \mathbf r_{00} + \mathbf v_0 t \\
\mathbf r_1 = \mathbf r_{10} + \mathbf v_1 t \\
\mathbf r = \mathbf r_{00} + \mathbf v t
\end{array}
$$
At $t = \Delta T$ we let $\mathbf r = \mathbf r_1$ and we compute $\mathbf v$ as $\frac 1{\Delta T} \left(\mathbf r_{10} - \mathbf r_{00}\right) + \mathbf v_1$. The point $P$ (i.e. the spacecraft) is then moving, in $\Delta T$, from $P_1$ to $P_2$. The necessary velocity increments to match the asteroid velocities are then:
\begin{equation}
\label{eq:dvs}
\begin{array}{l}
\Delta V_0 = \frac 1{\Delta T} \Delta \mathbf r + \Delta \mathbf v \\
\Delta V_1 =  \frac 1{\Delta T} \Delta \mathbf r
\end{array}
\end{equation}
The quantity $d_o = \sqrt{|\Delta \mathbf V_0|^2 + |\Delta \mathbf V_1|^2}$ is here proposed as a phasing indicator. In order to highlight that this indicator accounts for a linearized orbital geometry, we refer to it as the orbital indicator. The great thing about the orbital indicator is that if we associate in $t_0$ each asteroid $A_i$ to a vector defined as  $\mathbf x_i = [\frac{1}{\Delta T} \mathbf r_i + \mathbf v_i, \frac{1}{\Delta T} \mathbf r_i]$, then the orbital phasing indicator for the $A_i$, $A_j$ transfer, is simply the euclidean distance between the corresponding vectors $\mathbf x_i$, $\mathbf x_j$.

\subsubsection{Phasing indicators as phasing value surrogates}
\label{sec:knn}
Both phasing indicators introduced result in a fast ranking of transfer opportunities. Assume to have $t_0$, $\Delta T$ and $A_s$ (i.e. $S$ is sitting on an asteroid at $t_0$) and to have to rank all asteroids in  $\mathcal A$ as to consider only the first $k$ for a detailed computation of the orbital transfer. If one is to use the euclidean indicator, this task is efficiently solved by computing the $k$-nearest neighbours ($k$-NN) in $\mathcal A$ to $A_s$ using $\mathbf x = [\mathbf r, \mathbf v]$ (i.e. the asteroids position and velocities at $t_0$) to define the points in a six dimensional space. 
In a similar way, if one is to use the orbital indicator, the same task is as efficiently solved by computing the $k$-nearest neighbours using $\mathbf x = [\frac{1}{\Delta T} \mathbf r + \mathbf v, \frac{1}{\Delta T} \mathbf r]$ to define the points in a six dimensional space. In both cases, given the low dimensionality of the $k$-NN problem, a k-d tree data structure \cite{Bentley:1975:MBS:361002.361007} is an efficient choice to perform the computation. The complexity to build a static k-d tree is $O(N\log N)$, while the $k$-NN query has complexity $O(k\log N)$. One single $k$-NN computation including the construction of the k-d tree, on our test case, takes on average 0.25 seconds, while the computation of all the phasing values $\varphi$ to then extract the best $k$ targets, takes, on average, 5 minutes (tests made on an Intel(R) Core(TM) i7-4600U CPU having the clock at 3.3 GHz and with a cache of 4096 KB, exact implementation details available on-line as part of PyKEP.phasing module).

\begin{figure}
\centering
\includegraphics[width=0.95\columnwidth]{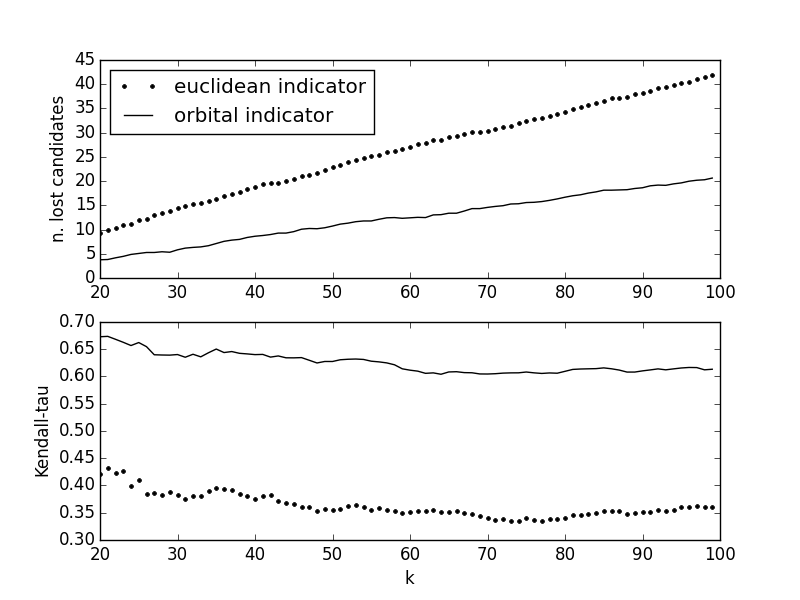}
\caption{Rank correlations of the proposed phasing indicators (average over 100 random cases)}
\label{fig:ranks_correlations}
\end{figure}

Such a speed increase (three orders of magnitude) is only useful if the resulting rankings are correlated. In Figure \ref{fig:ranks_correlations} we show the rank correlations between the ground truth rank, computed using $\varphi$ and those resulting from the newly introduced phasing indicators. The plot shows the average over 100 randomly selected $t_0$ and $A_s$. The value of the Kendall-tau coefficient is reported together with the number of false negatives, that is the asteroids that are within the best $k$ according to the $\varphi$ value, but are not within the $k$-NN computed using $d_e$ or $d_o$. 

The Kendall-tau coefficient is defined as $\tau = \frac{n_c - n_d}{(1 / 2 k(k-1))}$, where $n_c$ is the number of concordant pairs, whereas $n_d$ is the number of discordant pairs. A value of $\tau = 1$ corresponds two two identical rankings, similarly a $\tau = -1$ corresponds to two perfectly discordant rankings. In general, if the two rankings are uncorrelated, a value $\tau = 0$ is expected. The results show how both the new introduced quantities $d_e$ and $d_o$ are directly correlated with the the phasing value $\varphi$ and thus can be used as surrogates for the phasing value $\varphi$. The orbital indicator outperforms the euclidean indicator resulting in ranks better correlated with respect to the ground truth.

\subsection{Clustering asteroids}
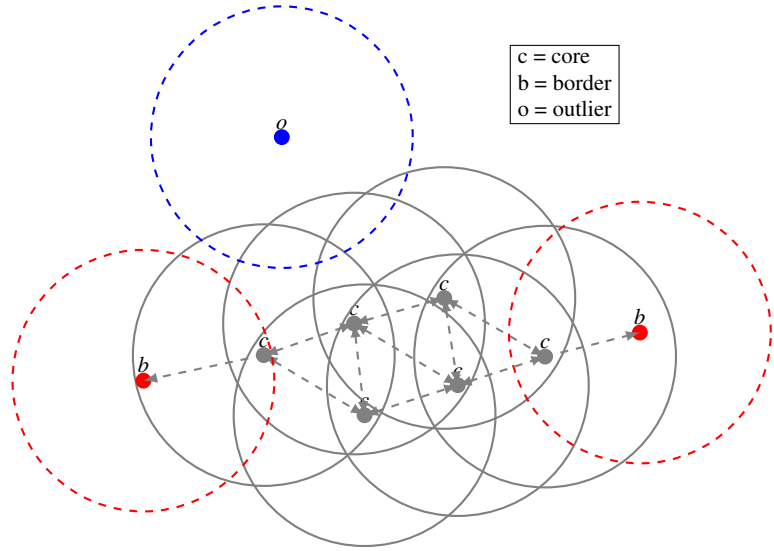
\begin{figure}
\centering
\begin{tikzpicture}[>=latex,scale=2.0]
 
  \pgfmathsetmacro{\dim}{1.5}
 
 \foreach \x in {(0,0),(0.6,0.21),(1.2,0.38),(0.67,-0.4),(1.29,-0.2),(1.87,-.01)} {
    \draw[thick, gray] \x circle(.87);
    \fill [gray] \x circle (\dim pt) ;
    \draw \x node[above] {$c$} ;
  }

  \draw[<->,thick,dashed, gray] (0,0) -- (0.6,0.21);
  \draw[<->,thick,dashed, gray] (0,0) -- (0.67,-0.4);
  \draw[<->,thick,dashed, gray] (0.6,0.21) -- (0.67,-0.4);
  \draw[<->,thick,dashed, gray] (0.6,0.21) -- (1.2,0.38);
  \draw[<->,thick,dashed, gray] (0.6,0.21) -- (1.29,-0.2);
  \draw[<->,thick,dashed, gray] (0.67,-0.4) -- (1.29,-0.2);
  \draw[<->,thick,dashed, gray] (1.2,0.38) -- (1.87,-.01);
  \draw[<->,thick,dashed, gray] (1.29,-0.2) -- (1.87,-.01);
  \draw[<->,thick,dashed, gray] (1.2,0.38) -- (1.29,-0.2);

  \foreach \x in {(-0.8,-0.17), (2.5,0.15)} {
    \draw[thick, red, dashed] \x circle(.87);
    \fill [red] \x circle (\dim pt) ;
    \draw \x node[above] {$b$} ;
  }
  
  \draw[->,thick,dashed, gray] (0,0) -- (-0.8,-0.17);
  \draw[->,thick,dashed, gray] (1.87,-.01) -- (2.5,0.15);
  
  \draw[thick, blue, dashed] (0.12,1.45) circle(.87);
  \fill [blue] (0.12,1.45) circle (\dim pt) ;
  \draw (0.12,1.45) node[above] {$o$} ;
  
  \node[draw,align=left] at (2, 1.8) {c = core\\ b = border\\ o = outlier};
\end{tikzpicture}
\caption{DBSCAN clustering illustration\label{fig:dbscan} for a 2-D case and a naive metric. $m_{pts}=3$.}
\end{figure}

Besides ranking possible transfer opportunities, the phasing value indicators $d_e$ and $d_o$ can be useful to define a metric over the set of all asteroids $\mathcal A$ at each $t_0$. Such a metric can then be used to compute, for a given $t_0$, asteroid clusters. Using the orbital metric $d_o$ as explained above, we define the points $\mathbf x_i = [\frac{1}{\Delta T} \mathbf r_i + \mathbf v_i, \frac{1}{\Delta T} \mathbf r_i]$ 
and apply clustering algorithms \cite{grabmeier2002techniques} directly on them. Large clusters of well-phased asteroids are likely to result in good opportunities for a multiple asteroid rendezvous mission. The clustering algorithm DBSCAN \cite{ester1996density} is particularly suitable to find clusters in this domain. The algorithm has two fundamental parameters: $\epsilon$, indicating the radius of the ball that defines each point neighbourhood and $m_{pts}$, defining the minimum number of neighbours necessary to be part of a cluster core. According to DBSCAN, an asteroid $A$ belongs to a cluster $\mathcal C$ (i.e. a subset of $\mathcal A$) if it either has at least $m_{pts}$ asteroids inside its $\epsilon$ neighbourhood or at least one of its neighbours does. In the first case $A$ is said to be in the cluster core, otherwise it is labelled as a border point. If the orbital metric is used, the $\epsilon$  neighbourhood has an interesting interpretation. Asteroids within the $\epsilon$ neighbourhood of $A$ will be reachable from $A$, according to the simple linear trajectory transfer model, with a transfer requiring a $\Delta V \le \epsilon$ and a transfer time of $T$. In Figure \ref{fig:dbscan} a simple example visualizing a cluster, as defined by DBSCAN, is shown. Asteroids that are not associated to any cluster are labelled as outliers. 

\begin{figure}
\centering
\includegraphics[width=.95\columnwidth]{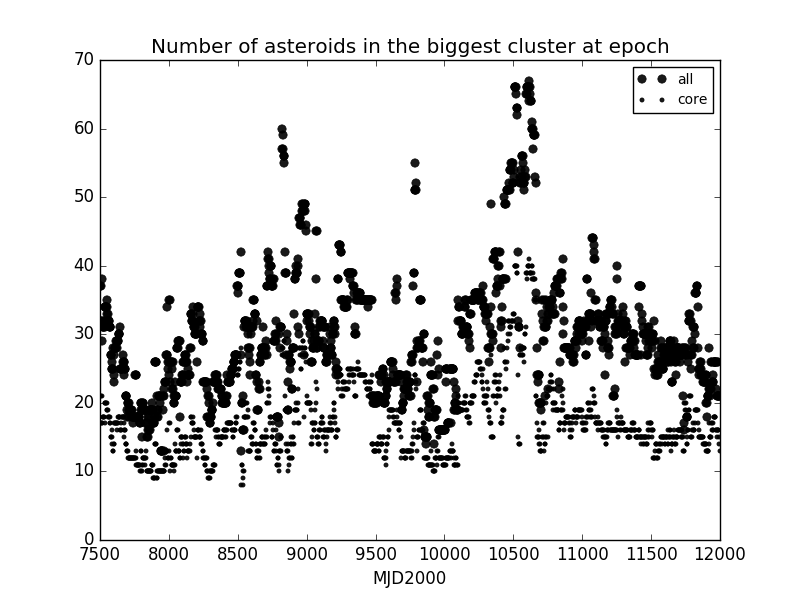}
\caption{Size of the largest cluster at epoch.\label{fig:cluster_size_vs_epoch}}
\end{figure}

Consider now our data-set and a starting epoch in a 3 days resolution grid defined in $[7500, 12000]$ [mjd2000]. At each epoch, one can run DBSCAN and compute all asteroid clusters setting $\epsilon = 1650$ [m/s] and $m_{pts} = 5$. The result is visualized in Figure \ref{fig:cluster_size_vs_epoch} where, at each epoch, the size of the largest cluster found is reported as well as the number of core asteroids in it. Such a graph is proposed as a tool to help selecting the target area in the main belt and the time frame of a possible multiple rendezvous mission. In this particular case, for example, one notes that around the MJD2000 $10500$ a special conjunction happens and relatively large clusters appear, having a size which is, on average, twice as much as clusters at other epochs. It comes then naturally that if a spacecraft was to be operating in one of these big cluster, it would have greater chances to find good transfer opportunities.

\begin{figure}
\centering
\includegraphics[width=0.80\columnwidth]{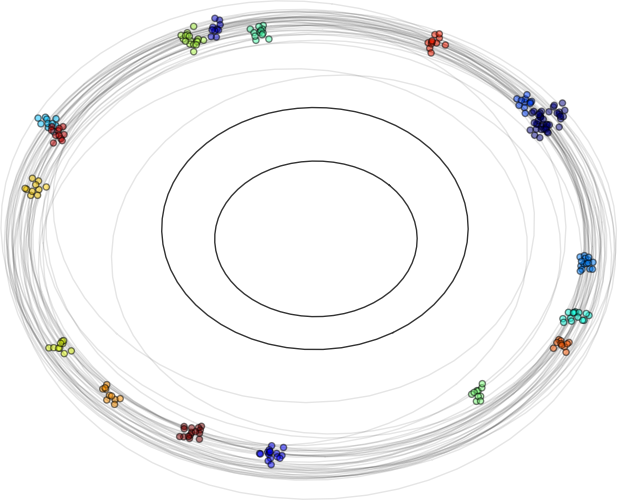}
\caption{Visualization of all clusters found by DBSCAN at $t_0 = 10614$ MJD2000. The orbital metric is used. CLusters show in different colors. The Earth and Mars orbits are also shown as reference.\label{fig:cluster_visualization}}
\end{figure}

\begin{figure}
\centering
\includegraphics[width=.95\columnwidth]{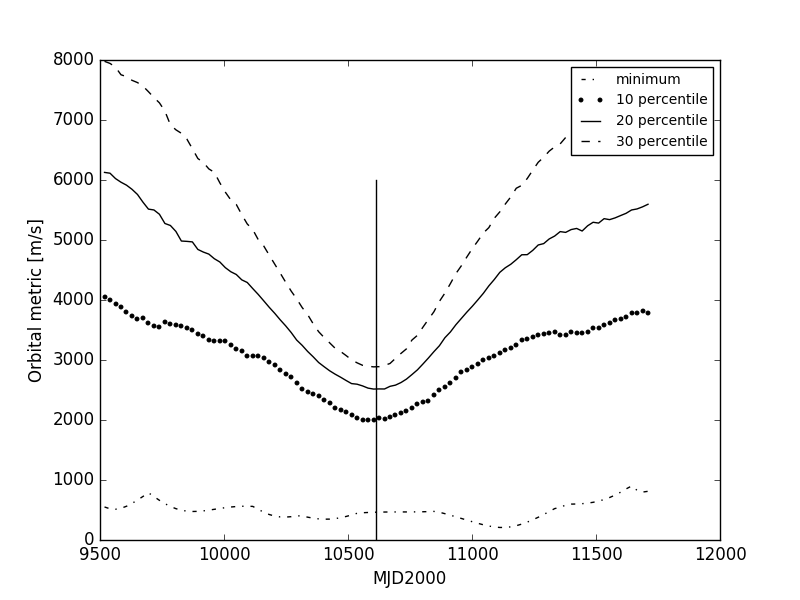}
\caption{Time evolution of the orbital metric computed for a cluster detected at $t_0= 10600$ [MJD2000]. All asteroid pairs are considered and an average over the best considered percentile is reported.
\label{fig:percentiles}}
\end{figure}

In Figure \ref{fig:cluster_visualization} the actual clusters computed for $t_0 = 10500$ MJD2000 are visualized together with some of the orbits of the main belt objects belonging to the data-set, the Earth and Jupiter orbit. The big cluster found by DBSCAN is clearly visible. All points in the cluster are guaranteed, in a first linear approximation, to be reachable from neighbours with a convenient orbital transfer having $\Delta V \le \epsilon = 1650$ [m/s]. If they are core members of the cluster they are guaranteed to be reachable from at least $m_{pts}$ other asteroids ($m_{pts} = 5$ in our case). Asteroids belonging to a cluster at $t_0$ are unlikely to form a cluster at different epochs as their orbital movement will tend to tear the cluster apart. Such an effect is directly proportional to $\epsilon$ as small values of $\epsilon$ imply similar orbital parameters. 

Asteroid clusters are defined at a given epoch $t_0$ and, due to orbital motion they may disperse more or less quickly. To show how, in this case, a cluster persists in time for some years, in Figure \ref{fig:percentiles} the time evolution of a particular cluster is analyzed over a six-year time span. The cluster we analyze is the biggest one detected by DBSCAN and can be spotted in Figure \ref{fig:cluster_visualization} as one big aggregate of points. A six-year window is defined being centered at the epoch $t_0 = 10614$ MJD2000 (when the cluster is detected). The orbital metric $d_o$ is then computed for all pairs of asteroids in the cluster at each epoch in the defined window, and the minimum, the 10, 20 and 30 percentiles are reported. The plot shows how the cluster is resilient to being disrupted, at least within the considered time window. Since the minimum of the orbital metric $d_o$ remains constantly low, at least two asteroids belonging to the cluster are always connected by an extremely advantageous orbital transfer, further more, while all the percentile plots present a minimum at the cluster epoch, they are only mildly increasing away from the cluster point. Similar features can be observed consistently for all clusters detected in the asteroid main belt.

\begin{algorithm}[t]
\caption{MOBS algorithm}\label{euclid}
\begin{algorithmic}[1]
\Procedure{MOBS}{$t_0\in [\overline t_0, \underline t_0], A_0\in\mathcal A$}
   \State $\mathcal B = \{[[A_0], 1, 1 ]\}$, best $= [[A_0], 1, 1 ]$ \Comment{Both resources are fully available at the beginning}
   \While{$\mathcal B \ne \emptyset $}
      \State $\mathcal L = \emptyset $
      \For{\textbf{each} $\mathcal N \in \mathcal B$}
      \State $\mathcal L = \mathcal L \cup$ Branch($\mathcal N$) \Comment{Branch() creates maximum $BF$ new nodes}
      \EndFor
      \State best $=$ UpdateBest($\mathcal L$)
      \State $\mathcal B = $Beam($\mathcal L$)  \Comment{Beam() selects maximum $BS$ nodes}
   \EndWhile\label{euclidendwhile}
   \State \textbf{return} best
\EndProcedure
\end{algorithmic}
\label{alg:bs}
\end{algorithm}

\subsection{Multi-objective beam-search}
\label{sec:mobs}
Having developed the phasing value and asteroid clustering methods, we are now ready to build a procedure to search for complete multiple asteroid rendezvous missions. We approach the problem, as formally stated in Sect. \ref{sec:problem_definition}, as a tree search problem. An algorithm based on the beam-search strategy is proposed, and its pseudo-code is shown as Algorithm \ref{alg:bs}. The algorithm, named MOBS, is a Multi-Objective Beam Search that accepts as inputs a starting epoch $t_0$ and a starting asteroid $A_0$ and searches for multiple rendezvous missions possible with the available resources. In MOBS, the non dimensional remaining mass $r_m = \frac{m-m_s}{m_p}$ and the non dimensional remaining time $r_t = 1- \frac{t-t_0}{tof}$ are considered as the two resources available to the spacecraft. A node, representing a multiple-rendezvous trajectory, is defined as a triplet containing a list of visited asteroids $[A_0, A_1, A_2, ... A_n]$, and the two remaining resources $r_t$ and $r_m$. The key elements of the algorithm are the Branch procedure and the Beam procedure. The Branch procedure is used to create branches from any particular node, which is equivalent to compute new transfers to one or more asteroids, increasing the overall objective of visiting as many asteroids as possible before the end of the mission. The Beam procedure is then used to select the most promising trajectories to be branched.

\subsubsection{Branch procedure}
The Branch procedure takes a node (that is a multiple rendezvous trajectory), selects a maximum of $BF$ (branching factor) possible asteroid targets and returns a list of new nodes created adding one of the target asteroids to the list of visited asteroids and updating $r_t$ and $r_m$ accordingly. The $BF$ asteroids are selected among the ones having the largest (i.e. best) phasing value. A phasing value surrogate is used to speed up the computation, so that the $BF$ asteroids will be the ones having the closest distance in the orbital metric $d_o$. Bodies already belonging to the list of visited asteroids are excluded as targets. The use of a k-d tree data structure makes the k-NN ($k = BF$) computations very efficient as mentioned in Sect. \ref{sec:knn}. For each of the target asteroids selected, the minimum arrival epoch optimal control problem is then solved first. If a solution is found, $t^*$ being the earliest epoch the spacecraft can reach the target asteroid, also fixed arrival time, minimum mass optimal transfers are computed with $t_f = t^* + i dt$, $i = 1 .. \ell$. For each feasible solution thus found, one can then compute the remaining resources $r_t$ and $r_m$ and insert a new node to the branched trajectory list. This branching procedure will return at most $\ell \cdot BF$ new trajectories visiting one more asteroid with respect to the parent node.

\subsubsection{Beam procedure}

\begin{table}[t]
\begin{center}
\caption{Node value definitions}
\label{tab:node_value_defs}
\begin{tabular}{p{2cm}p{2.4cm}p{2cm}p{4.9cm}}
\hline\noalign{\smallskip}
$r_1=r_m$ & best mass\\
$r_2=r_t$ & best time\\
$r_3=\frac{1}{2}r_m+\frac{1}{2}r_t$ & average\\
$r_4=\frac{r_m r_t}{r_m+r_t}$ & soft min\\
$r_5=\min(r_m,r_t)$ & hard min\\
$r_6$ & Pareto dominance\\
\noalign{\smallskip}\hline\noalign{\smallskip}
\end{tabular}
\end{center}
\end{table}

\begin{figure}[t]
\centering
\subfloat[$r_1$: best mass]{
    \includegraphics[viewport=14 12 566 386,clip,width=0.49\columnwidth]{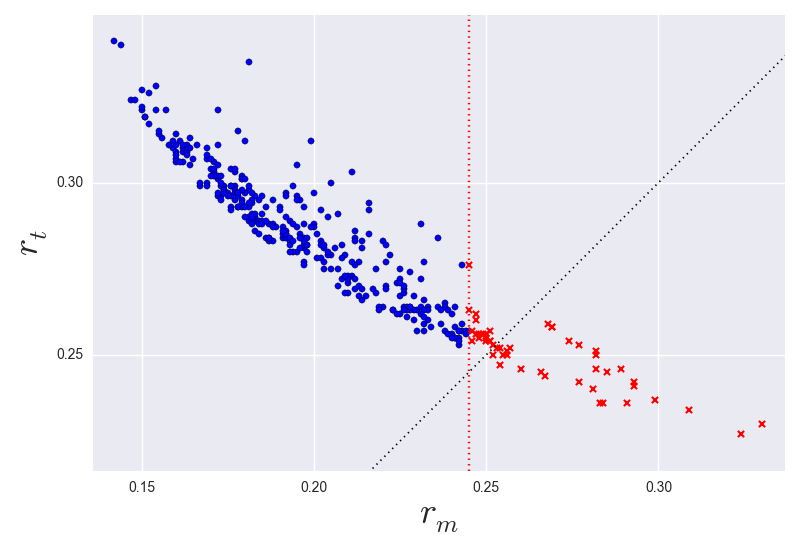}}
\subfloat[$r_2$: best time]{
    \includegraphics[viewport=14 12 566 386,clip,width=0.49\columnwidth]{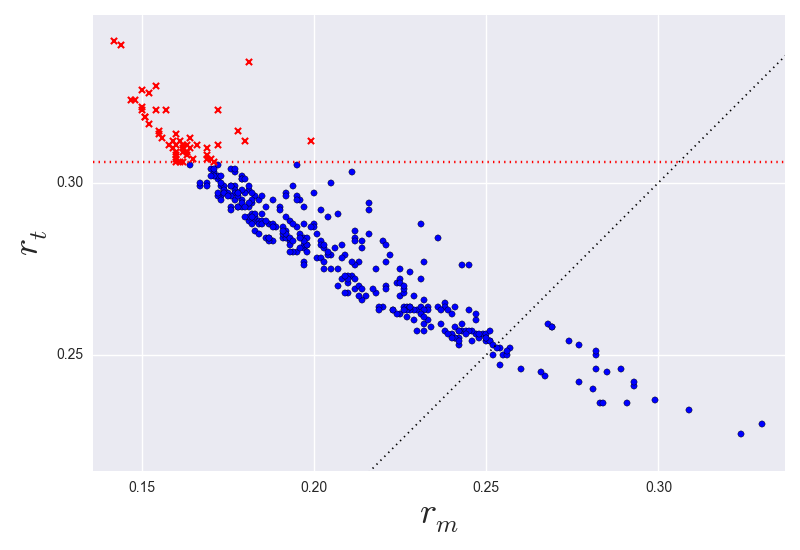}}\\
\subfloat[$r_6$: Pareto dominance]{
    \includegraphics[viewport=14 12 566 386,clip,width=0.49\columnwidth]{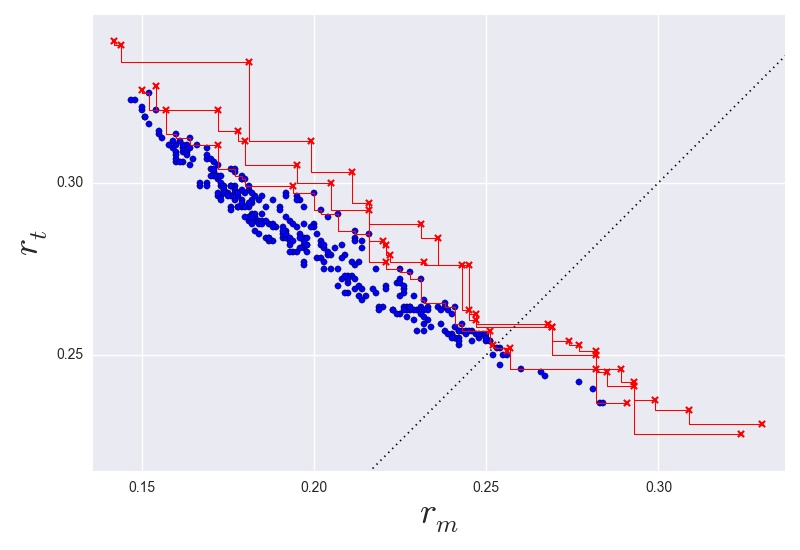}}
\subfloat[$r_3$: average]{
    \includegraphics[viewport=14 12 566 386,clip,width=0.49\columnwidth]{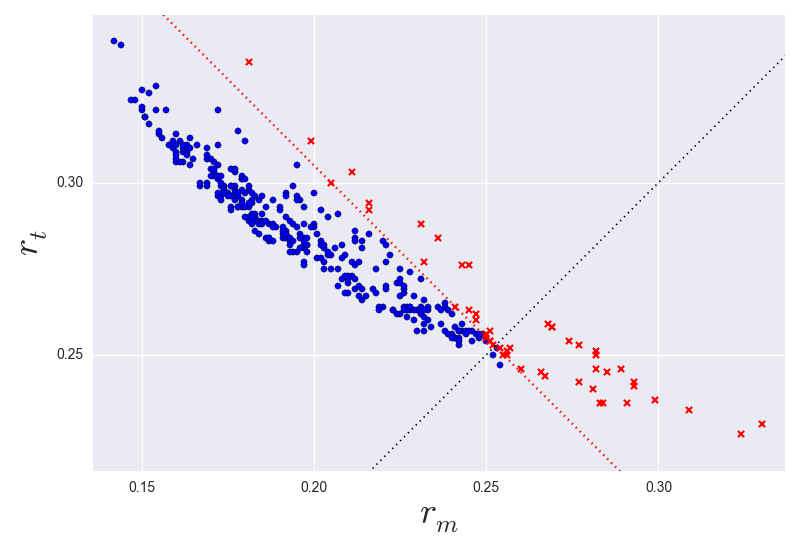}}\\
\subfloat[$r_4$: soft min]{
    \includegraphics[viewport=14 12 566 386,clip,width=0.49\columnwidth]{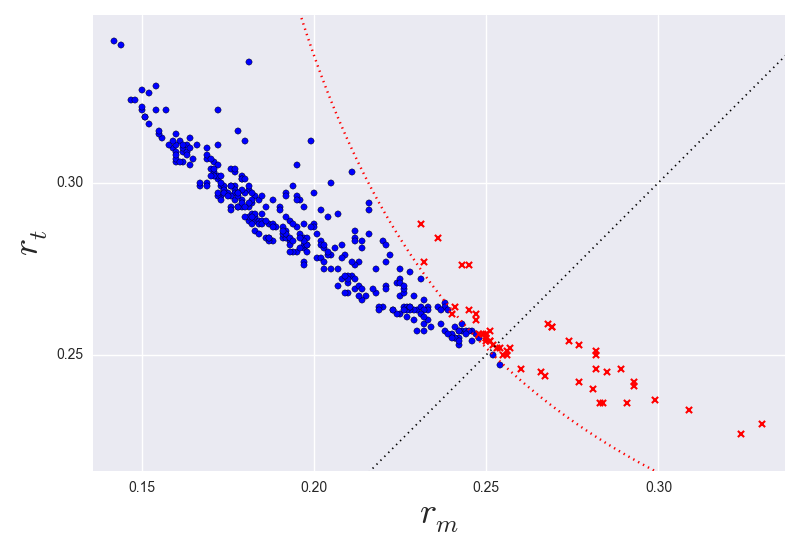}}
\subfloat[$r_5$: hard min]{
    \includegraphics[viewport=14 12 566 386,clip,width=0.49\columnwidth]{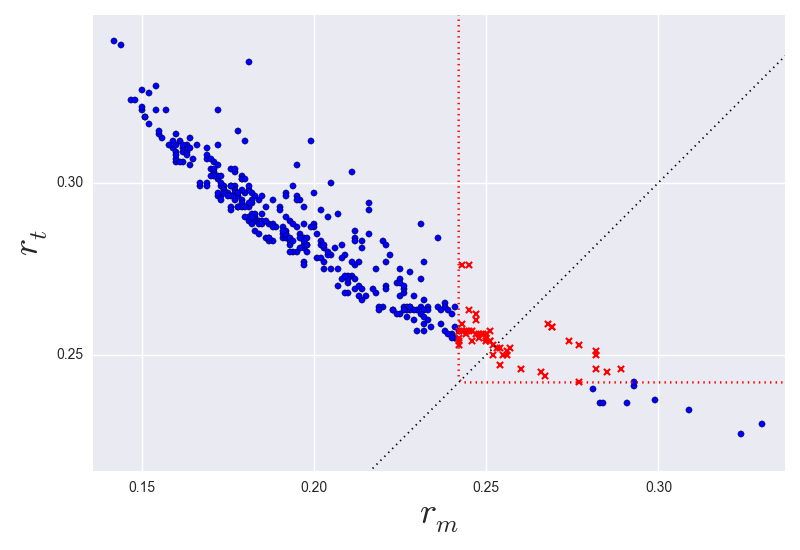}}
\caption{Possible rankings for beam search\label{fig:node_selection}: top 50 nodes at a given tree's current depth, as determined by the rating functions in Table \ref{tab:node_value_defs}.}
\end{figure}

The Beam procedure takes a set of nodes and returns one of its subsets having at maximum $BS$ members. In other words, it selects from a list of multiple rendezvous trajectories having equal depth, the most promising $BS$ to be carried forward in the search. This selection is crucial to the performance of the overall scheme and is made introducing a node value $r$, computed for each node. The best $BS$ nodes with respect to this value are returned. A trajectory should be considered good when it made clever use of the spacecraft's available resources, thus both $r_m$ and $r_t$ should be considered in defining the node value $r$. The first trivial choice would be to use directly $r_m$ or $r_t$ as a definition for the node value. This way, trajectories having spent a minimum amount of propellant, or time would be considered for further expansion. Since the phasing value is used to branch nodes, trajectories in the list already have been indirectly pre-selected with respect to a multi-objective criteria (the phasing value is defined with respect to the hypervolume), thus such a trivial choice would be less greedy than it appears and in fact, it works reasonably well when one knows upfront that one of the two available resources is particularly scarce. In the general case, though, a multi-objective aggregation of the two objectives seems like a more promising option to directly select good candidate trajectories. A number of options are thus proposed and summarized in Table \ref{tab:node_value_defs}. A first direct approach is to consider a node value aggregating the two resources into one number via the expression $r  = \lambda_m r_m + \lambda_t r_t$. The weights $\lambda_m$ and $\lambda_t$ implicitly define a priority on the two resources. One could thus consider them to be equal, in which case a simple \emph{average} node rank is defined. The average node rank allows for a node having one of the resources almost completely depleted and the other fully available to rank equally to a node having both resources consumed half-way, which does not seem as a good choice. This suggests to introduce the \emph{soft min} \cite{gtoc5} case where the weights $\lambda$ are adaptively modified according to how much a certain trajectory has used of a certain objective via the expression $\lambda_j = \left(1 - \frac{r_j}{r_m+r_t}\right)$. It is easy to recognize that this adaptive weight scheme is equivalent to use $r = \frac{r_m r_t}{r_m+r_t}$ for ranking. One possible problem with the \emph{soft min} approach is that the weights, though adaptive, are still implicitly defining the importance of the two resources (mass and time) in a somewhat arbitrary fashion.
A different approach is to use Pareto dominance concepts,
where the top nodes are determined through a combination of non-dominated sorting and usage of a \emph{Crowded Comparison} operator \cite{deb2002fast}.

\section{Experiments}
We evaluate the overall performance of MOBS to search for multiple asteroid rendezvous missions. We consider the GTOC7 data so that the asteroids are moving along well defined Keplerian orbits. We also set a minimum waiting time on the asteroid $t_w = 30$ [days], an initial spacecraft mass $m_0 = 2000$ [kg], an initial propellant mass $m_p = 1200$ [kg], a maximum thrust $\tau_M = 0.3$ [N] and a specific impulse $I_{sp} = 3000$ [sec.]. We consider a maximum total mission duration of $tof = 6$ [years] and a starting epoch $t_0 \in [7500, 12000]$ [mjd2000]. These values create a well defined instance of the multiple asteroid rendezvous problem defined in Sect. \ref{sec:problem_definition}.

\begin{table}[t]
\begin{center}
\caption{Multi-Objective Beam Search (MOBS) performance}
\label{tbl:mobs_performance}
\begin{tabular}{ccrrrrrrr}
\hline\noalign{\smallskip}
 & node value  & $\geq$8 & $\geq$9 & $\geq$10 & $\geq$11 & $\geq$12 & $\geq$13 & $\geq$14 \\
\noalign{\smallskip}\svhline\noalign{\smallskip}

\multirow{4}{*}{with cluster} & mass        & 79.4 & 10.2 &  0.0 &  0.0 &  0.0 &  0.0 &  0.0 \\ 
& time        & 95.0 & 95.0 & 87.2 & 57.6 & 23.2 &  3.8 &  0.0 \\ 
& Pareto      & 94.8 & 93.2 & 70.2 & 33.4 &  7.4 &  0.6 &  0.0 \\ 
& soft min    & 95.0 & 94.8 & 83.0 & 47.2 & 10.0 &  0.6 &  0.0 \\ 
\hline
\multirow{4}{*}{without cluster} & mass        & 67.4 &  6.0 &  0.2 &  0.0 &  0.0 &  0.0 &  0.0 \\ 
& time        & 93.6 & 92.6 & 86.0 & 49.0 & 12.2 &  0.6 &  0.0 \\ 
& Pareto      & 93.4 & 91.0 & 64.0 & 22.4 &  2.2 &  0.2 &  0.0 \\ 
& soft min    & 93.6 & 92.8 & 79.0 & 35.2 &  5.4 &  0.2 &  0.0 \\

\noalign{\smallskip}\hline\noalign{\smallskip}
\end{tabular}
\end{center}
\end{table}

\begin{figure}
\centering
\includegraphics[width=0.95\columnwidth]{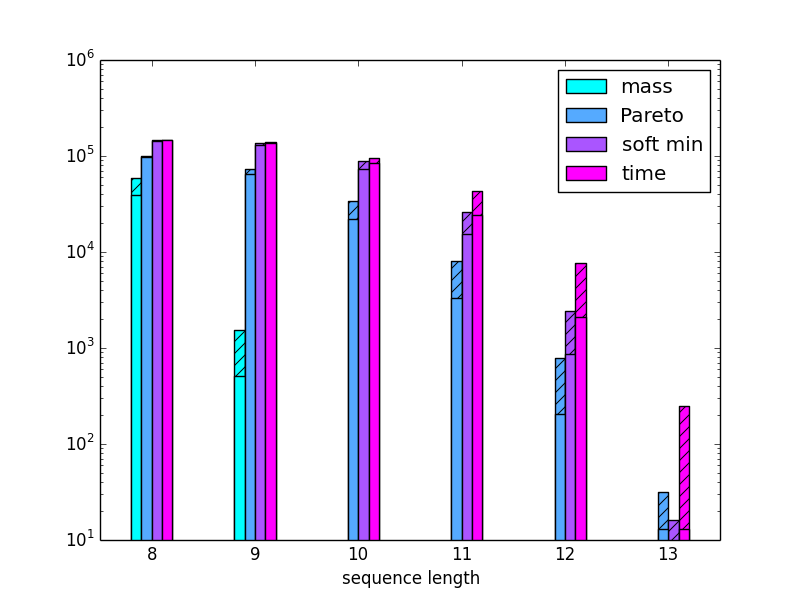}
\caption{Number of sequences generated by 500 searches for each node value estimator. The hatched part corresponds to the starting condition with clustering. \label{fig:mobs}}
\end{figure}

Table~\ref{tab:node_value_defs} shows the node value estimates used to evaluate MOBS. 
For all experiments we set $BF$ to 10 and use the orbital metric for clustering as well as $k$-NN search. Additionally, MOBS requires a starting epoch $t_0$ and a starting asteroid $A_0$.
We evaluate two different ways to provide such an initial condition: 
\begin{itemize}
\item sampling $t_0$ uniformly from the launch window and choosing a random asteroid $A_0$ from the set of all 16,256 asteroids
\item sampling $t_0$ uniformly from the launch window; perform clustering at epoch $t_0$ and select $A_0$ from the core of the largest cluster.
\end{itemize}
For each initial condition and node value estimator $r \in \{ r_1, r_2, r_4, r_6\}$, we run $500$ tree searches. A search results in a number of solutions of which we record the longest sequence. Table~\ref{tbl:mobs_performance} shows the percentage of MOBS runs that resulted in a sequence of length at least $8, 9, \ldots, 14$ asteroids. The total number of solutions that reached a given length is reported in Figure~\ref{fig:mobs}. On average, one full MOBS run took one hour so that the entire experimental campaign here reported, involving 4000 tree searches, was run on a machine having 20 cores at 3.1GHZ (40 parallel hyper-threads) during a period of, roughly, 4 days. During the runs, an approximate number of 30,000,000 OCPs are solved.
\begin{figure}
\centering
\includegraphics[width=0.9\columnwidth]{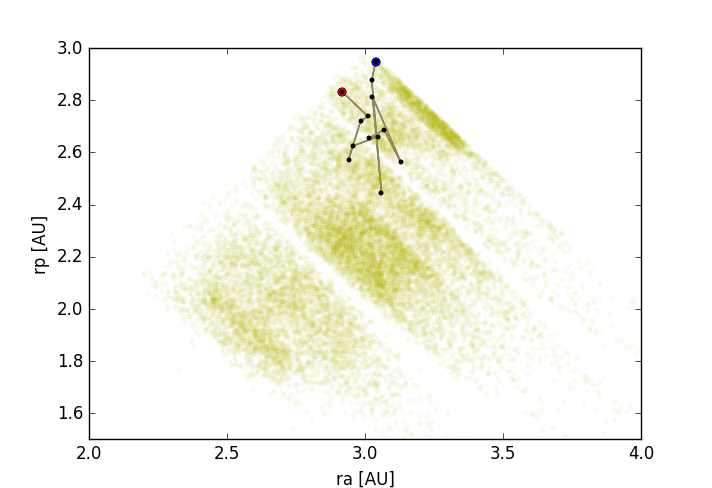}
\includegraphics[width=0.9\columnwidth]{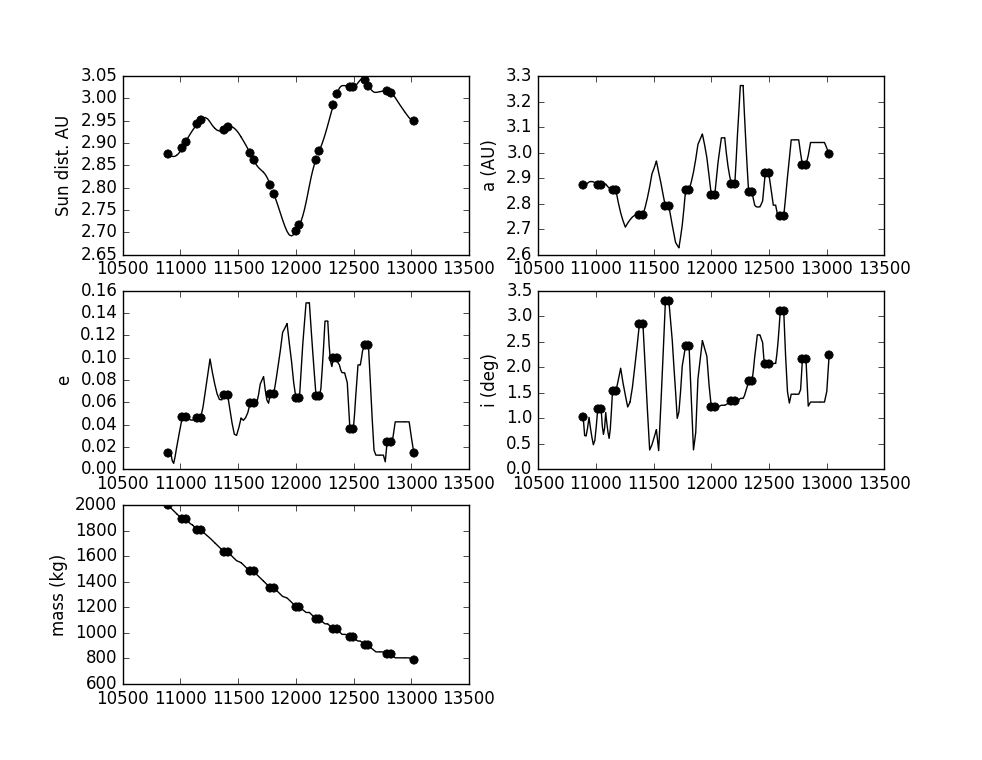}
\caption{Visualization of a multiple asteroid rendezvous mission visiting 13 bodies. Top: Aphelion and perihelion of the 13 asteroids visited in one of the mission designed by MOBS. The sequence starts at the blue dot. Asteroids from the main-belt population are also shown as background for reference. Bottom: some details of one of the trajectories found by MOBS. Each dot corresponds to a departure or arrival at one of the asteroids. Orbital parameters are the Keplerian osculating parameters. \label{fig:orbital_param}}
\end{figure}

The maximum asteroid sequence length reported by participants of the GTOC7 as part of their final solutions was 13 (as reported during the GTOC7 workshop hosted in Rome in May, 2015). It is still an open question whether a sequence of length 14 exists, under the given settings, though it appears plausible. The proposed MOBS algorithm is able to find sequences of length 12 with ease and it does find, in a small percentage of runs, longer sequences of asteroids to visit. MOBS never returned, neither during the experiments nor during the original runs made within the competition time-frame, a valid sequence of length 14. 

Table~\ref{tbl:mobs_performance} shows that the best node value estimator is time, while using mass result in extremely poor performances. While difficult to know upfront, this implies that the mission resources $r_m$ and $r_t$ are not evenly balanced: the spacecraft has comparatively more propellant than mission time in order to achieve its goals. Under different initial conditions, propellant could be a resource as scarce as time in which case other indicators should deliver better trajectories by exploiting the multi-objective trade-offs. The use of asteroid clustering greatly improves the chances to find good opportunities, focusing the search in the most promising areas of the asteroid belt at a given epoch. 

We conclude this book chapter reporting, in Figure \ref{fig:orbital_param}, some visual information on one of the missions designed by MOBS and visiting 13 asteroids. We show the perihelion and aphelion for each of the asteroids visited overlapped to the background population of asteroids considered. We also show the osculating Keplerian elements during the entire mission as well as the spacecraft mass. The mission operates in the outer part of the main belt, acquiring a minimum distance of roughly 2.4 AUs and a maximum of 2.9 AUs. From the rapid increases and drops of the osculating semi-major axis during the same asteroid to asteroid transfers, a lot of thrust is used to cope with the non-perfect phasing.
A control strategy that, though clearly sub-optimal when considering a single leg alone, allows to visit a greater number of asteroids when adopted to assemble the whole sequence.

\section{Conclusions}
The design of complex interplanetary trajectories is the subject of the Global Trajectory Optimization Competitions. Participating in these events requires a solid knowledge of basic astronomical problems such as the Kepler's problem, the Lambert's problem, the perturbed Kepler's problem, a certain familiarity with optimal control theory and algorithms as well as the development of original and innovative methods tailored for the particular problem to be solved. In the case of the 7th edition of this competition, the possibility to design multiple asteroid rendezvous missions was part of the problem assigned. We have found that a phasing value, defined as the hypervolume of the Pareto front of the multi-objective Lambert's transfer can be conveniently introduced and approximated using a surrogate orbital indicator. The phasing value approximation can be used to rank possible transfer opportunities and as a metric to define asteroid clusters in the main asteroid belt. We use these ideas to assemble a multi-objective tree search able to consistently design, in the GTOC7 data set, multiple rendezvous missions visiting up to 13 asteroids.

\begin{acknowledgement}
Lu{\'i}s F. Sim{\~o}es was supported by FCT (Minist\'erio da Ci\^encia e Tecnologia) Fellowship SFRH/BD/84381/2012.
\end{acknowledgement}

\bibliographystyle{spmpsci}
\bibliography{multastbib}

\begin{thebibliography}{10}
\providecommand{\url}[1]{{#1}}
\providecommand{\urlprefix}{URL }
\expandafter\ifx\csname urlstyle\endcsname\relax
  \providecommand{\doi}[1]{DOI~\discretionary{}{}{}#1}\else
  \providecommand{\doi}{DOI~\discretionary{}{}{}\begingroup
  \urlstyle{rm}\Url}\fi

\bibitem{bdack1991survey}
B\"ack, T., Hoffmeister, F., Schwefel, H.: A survey of evolution strategies.
\newblock In: Proceedings of the 4th international conference on genetic
  algorithms, pp. 2--9 (1991)

\bibitem{battin}
Battin, R.H.: An introduction to the mathematics and methods of astrodynamics.
\newblock AIAA (1999)

\bibitem{Bentley:1975:MBS:361002.361007}
Bentley, J.L.: Multidimensional binary search trees used for associative
  searching.
\newblock Commun. ACM \textbf{18}(9), 509--517 (1975)

\bibitem{worhp}
B{\"u}skens, C., Wassel, D.: The esa nlp solver worhp.
\newblock In: Modeling and Optimization in Space Engineering, pp. 85--110.
  Springer (2013)

\bibitem{gtoc7:problem}
Casalino, L., Colasurdo, G.: {Problem Description for the 7th Global Trajectory
  Optimisation Competition} (2014).
\newblock \urlprefix\url{http://areeweb.polito.it/gtoc/gtoc7_problem.pdf}.
\newblock [Online; accessed 10-March-2016]

\bibitem{chaslot2006monte}
Chaslot, G., Saito, J.T., Bouzy, B., Uiterwijk, J., Van Den~Herik, H.J.:
  Monte-carlo strategies for computer go.
\newblock In: Proceedings of the 18th BeNeLux Conference on Artificial
  Intelligence, Namur, Belgium, pp. 83--91. Citeseer (2006)

\bibitem{conway2010spacecraft}
Conway, B.A.: Spacecraft trajectory optimization, vol.~29.
\newblock Cambridge University Press (2010)

\bibitem{cormen2001c}
Cormen, T.H., Leiserson, C.E., Rivest, R.L.: C. stein introduction to
  algorithms.
\newblock MIT Press \textbf{5}(3), 55 (2001)

\bibitem{d2006apies}
D'Arrigo, P., Santandrea, S.: The {APIES} mission to explore the asteroid belt.
\newblock Advances in Space Research \textbf{38}(9), 2060--2067 (2006)

\bibitem{deb2002fast}
Deb, K., Pratap, A., Agarwal, S., Meyarivan, T.: A fast and elitist
  multiobjective genetic algorithm: {NSGA-II}.
\newblock Evolutionary Computation, IEEE Transactions on \textbf{6}(2),
  182--197 (2002)

\bibitem{ester1996density}
Ester, M., Kriegel, H.P., Sander, J., Xu, X.: A density-based algorithm for
  discovering clusters in large spatial databases with noise.
\newblock In: Kdd, vol.~96, pp. 226--231 (1996)

\bibitem{snopt}
Gill, P.E., Murray, W., Saunders, M.A.: Snopt: An sqp algorithm for large-scale
  constrained optimization.
\newblock SIAM journal on optimization \textbf{12}(4), 979--1006 (2002)

\bibitem{grabmeier2002techniques}
Grabmeier, J., Rudolph, A.: Techniques of cluster algorithms in data mining.
\newblock Data Mining and Knowledge Discovery \textbf{6}(4), 303--360 (2002)

\bibitem{mcts}
Hennes, D., Izzo, D.: Interplanetary trajectory planning with monte carlo tree
  search.
\newblock In: Proceedings of the 24th International Conference on Artificial
  Intelligence, pp. 769--775. AAAI Press (2015)

\bibitem{pykep}
Izzo, D.: {PyGMO and PyKEP: open source tools for massively parallel
  optimization in astrodynamics (the case of interplanetary trajectory
  optimization)}.
\newblock In: Proceed. Fifth International Conf. Astrodynam. Tools and
  Techniques, ICATT (2012)

\bibitem{izzolambert}
Izzo, D.: Revisiting lambert's problem.
\newblock Celestial Mechanics and Dynamical Astronomy \textbf{121}(1), 1--15
  (2014)

\bibitem{gtoc6}
Izzo, D., Sim{\~o}es, L.F., M\"artens, M., de~Croon, G.C.H.E., Heritier, A.,
  Yam, C.H.: {Search for a Grand Tour of the Jupiter Galilean Moons}.
\newblock In: {Proceedings of the 15th Annual Conference on Genetic and
  Evolutionary Computation (GECCO 2013)}, pp. 1301--1308. ACM Press, New York,
  NY, USA (2013)

\bibitem{gtoc5}
Izzo, D., Sim{\~o}es, L.F., Yam, C.H., Biscani, F., {Di Lorenzo}, D., Addis,
  B., Cassioli, A.: {GTOC5: Results from the European Space Agency and
  University of Florence}.
\newblock {Acta Futura} \textbf{8}, 45--55 (2014).
\newblock \doi{10.2420/AF08.2014.45}

\bibitem{taylor}
Jorba, {\`A}., Zou, M.: A software package for the numerical integration of
  odes by means of high-order taylor methods.
\newblock Experimental Mathematics \textbf{14}(1), 99--117 (2005)

\bibitem{NLP}
Kuhn, H.W.: Nonlinear programming: a historical view.
\newblock In: Traces and Emergence of Nonlinear Programming, pp. 393--414.
  Springer (2014)

\bibitem{messenger}
McAdams, J.V., Dunham, D.W., Farquhar, R.W., Taylor, A.H., Williams, B.G.:
  Trajectory design and maneuver strategy for the messenger mission to mercury.
\newblock Journal of spacecraft and rockets \textbf{43}(5), 1054--1064 (2006)

\bibitem{gtoc5jpl}
Petropoulos, A.E., Bonfiglio, E.P., Grebow, D.J., Lam, T., Parker, J.S.,
  Arrieta, J., Landau, D.F., Anderson, R.L., Gustafson, E.D., Whiffen, G.J.,
  Finlayson, P.A., Sims, J.A.: {GTOC5: Results from the Jet Propulsion
  Laboratory}.
\newblock {Acta Futura} \textbf{8}, 21--27 (2014).
\newblock \doi{10.2420/AF08.2014.21}

\bibitem{smart1}
Racca, G., Marini, A., Stagnaro, L., Van~Dooren, J., Di~Napoli, L., Foing, B.,
  Lumb, R., Volp, J., Brinkmann, J., Gr{\"u}nagel, R., et~al.: Smart-1 mission
  description and development status.
\newblock Planetary and space science \textbf{50}(14), 1323--1337 (2002)

\bibitem{sims}
Sims, J.A., Flanagan, S.N.: Preliminary design of low-thrust interplanetary
  missions.
\newblock In: AAS/AIAA Astrodynamics Specialist Conference, AAS Paper, pp.
  99--338 (1999)

\bibitem{vallado}
Vallado, D.A., McClain, W.D.: Fundamentals of astrodynamics and applications,
  vol.~12.
\newblock Springer Science \& Business Media (2001)

\bibitem{ipopt}
W{\"a}chter, A., Biegler, L.T.: On the implementation of an interior-point
  filter line-search algorithm for large-scale nonlinear programming.
\newblock Mathematical programming \textbf{106}(1), 25--57 (2006)

\bibitem{cassini}
Wolf, A.A.: Touring the saturnian system.
\newblock Space science reviews \textbf{104}(1-4), 101--128 (2002)

\bibitem{zhang2007moea}
Zhang, Q., Li, H.: {MOEA/D}: A multiobjective evolutionary algorithm based on
  decomposition.
\newblock Evolutionary Computation, IEEE Transactions on \textbf{11}(6),
  712--731 (2007)

\bibitem{zitzler1998multiobjective}
Zitzler, E., Thiele, L.: Multiobjective optimization using evolutionary
  algorithms—a comparative case study.
\newblock In: Parallel problem solving from nature--PPSN V, pp. 292--301.
  Springer (1998)

\end{thebibliography}

\end{document}